\documentclass[reprint,fleqn,superscriptaddress,twocolumn,showpacs,
amsmath,amssymb,prb,aps,bibliography]{revtex4-1}
\usepackage[all]{xy}
\usepackage{gensymb}
\usepackage{graphicx,psfrag,times,epsfig,color}
\usepackage{verbatim,natbib}

\usepackage{color}
\usepackage{makeidx}
\usepackage{amsmath}
\usepackage{bm}
\usepackage{amsfonts}
\usepackage{amssymb}
\usepackage{hyperref}

\begin{document}

\title{A framework for the description of charge order, pseudo and 
superconducting gap, critical 
temperature and pairing interaction in cuprate superconductors}
\author{E. V. L. de Mello}
\affiliation{Instituto de F\'{\i}sica, Universidade Federal Fluminense, 24210-346 Niter\'oi, RJ, Brazil}

\email[Corresponding author: ]{evandro@if.uff.br}

\begin{abstract}

 A unified phenomenological description framework is proposed for the evaluation of some of the
most important observables of the cuprate superconductors:
the pseudogap (PG) $\Delta_{\rm PG}$, the local superconducting amplitudes
$\Delta_{\rm SC}(r_i)$, the critical temperature $T_{\rm c}$ and 
charge ordering (CO) parameters. Recent detailed measurements of CO structures
and CO wavelengths $\lambda_{\rm CO}$ are faithfully
reproduced by solutions of a Cahn-Hilliard differential equation with a free energy 
potential $V_{\rm GL}$ that produces alternating small charge modulations.
The charge oscillations induce atomic
fluctuations that mediate the SC pair interaction proportional
to the $V_{\rm GL}$ amplitude. The local SC amplitude and phase
$\theta_i$ are connected by Josephson coupling $E_{\rm J}(r_{\rm ij})$
and the SC long-range order transition occurs when 
${\left < E_{\rm J} \right>} \sim k_{\rm B}T_{\rm c}$. 
The calculated results of the wavelength $\lambda_{\rm CO}$, $\Delta_{\rm PG}$, 
$\left < \Delta_{\rm SC}\right >$
and $T_{\rm c}$ calculations are in good agreement with a variety of experiments.
\end{abstract}
\maketitle 

A great deal of effort has been devoted to  the investigation of the 
different energy scales of high-temperature superconductors, their hole-doping 
dependence and, most importantly, their interconnections\cite{InterOrder2015}. 
 These might provide clues
to the pairing strength and to the superconducting (SC) 
mechanism. Under this program, Raman scattering on compounds 
with distinct average values of $p$ holes per Cu atoms\cite{Huefner2007,Raman2011}
identified vibration modes along
the nodal direction (B$_{2g}$) with energy $\Delta_{\rm c}(p)$ that follows 
closely $T_{\rm c}(p)$ and a second vibration mode measured along 
the antinode (B$_{1g}$), identified with $\Delta_{\rm PG}(p)$ because 
it correlates well with the PG temperature $T^*(p)$. Other
experiments like specific heat\cite{Tallon2003},
angle-resolved photon emission (ARPES)\cite{Yoshida2012,Yoshida2007},
scanning tunneling microscopy (STM)\cite{Kato2008,STM.Do2008} and submicron
Josephson junction tunneling\cite{Tunnel2012} identified also
$\Delta_{\rm PG}(p)$ but measured
another gap function ($\Delta_0(p)$) that increases slowly in the
underdoped region. 
In the overdoped region, $\Delta_0(p)$ stays close
to the PG and decreases rapidly beyond $p \sim 0.20$ holes/Cu or simply
$p \sim 0.20$. The interconnections
among these three energy scales and their roles in the phase diagram of cuprates
is the purpose of this letter. The present approach is 
complementary to that of Ref. \onlinecite{InterOrder2015} which studied several theoretical 
microscopic models and techniques with a $d$-wave SC order parameter 
and with  pair density wave (PDW) fields that lead to the 
concept of intertwined orders. 

To reveal this connection is of fundamental importance to consider also the
leading role of the ubiquitous spontaneous symmetry breaking or anomalous 
incommensurate charge-ordering (CO)\cite{InterOrder2015,Comin2016}. 
In particular, it was verified that
the CO wavelength is correlated with the distance between the Fermi arcs 
tips, establishing an intriguing connection between CO in real space and the 
PG in $k$-space on Bi$_2$Sr$_{2-{\rm x}}$La$_{\rm x}$O$_{6+\delta}$ (Bi2201) \cite{Wise2008,Comin2014}.
Many other experiments measure some kind of instability near $T^*(p)$, for instance, 
polar Kerr effect\cite{Kerr2008} and optical polarization rotation\cite{OptBirefri2014}.
On the other hand,
inhomogeneous magnetic-field response to muon spin rotation ($\mu$-SR)\cite{Muon2013}, STM\cite{Gomes2007,Parker2010}
and measurements of charge density wave (CDW) or CO by x-ray or resonant x-ray
scattering (REXS) 
\cite{Wise2008,Wu2011,Chang2012,Blanco-Canosa2014,Huecker2014,DaSilvaNeto2014,Campi2015,
Comin2014,Comin2015a,Comin2016,Wu2017,Tabis2017} 
have maximum signals near 
$p = 0.12$ and do not follow the increasing trend of $T^*$ when $p\rightarrow 0$,
probably because of the vanishing of the available charge. However,
all these observations may be regarded as distinct manifestations of an intrinsic
mesoscopic electronic phase separation with onset transition temperature
$T_{\rm PS}$ near $T^*(p)$\cite{Fradkin2012}, and this is a pillar 
of our approach. We recall, for further reference, that some
systems like La$_{\rm 2-x}$Sr$_{\rm x}$CuO$_4$ have their average doping level
x equal to the charge level $p$ or the average number of holes/Cu, while 
for others, these quantities are not equal but proportional.

Nanoscale electronic phase separations are predicted theoretically
by many different microscopic models,
mostly based on the Hubbard Hamiltonian, like for instance
Refs. [\onlinecite{Plakida2003,Plakida2016,Vojta2002,DiCastro2008,Kivelson2017,AMarie2010,InterOrder2015}]. 
These rigorous calculations are important to endorse the phenomenon of 
electronic phase separation on highly correlated systems like the cuprates,
but they neither reproduce the small variations\cite{Comin2016} of 
$\lambda_{\rm CO}(p)$
nor the very fine charge modulations $\Delta p \approx  10^{-2-3}$, like
in YBa$_2$Cu$_3$O$_{6+\delta}$ 
(Y123)\cite{Kharkov2016}.
Another important point is that charge density modulations are unambiguously present 
in the entire system at low temperature and even above $T_{\rm c}$ according to 
STM data\cite{Lang2002,Hanaguri2004,Gomes2007,Wise2008,Kato2008,Parker2010} and 
not in puddles occupying a volume fraction.
This last point appears to be in conflict with the finite CO correlation 
lengths\cite{Chang2012,Blanco-Canosa2014,Huecker2014,Comin2015a} but it is because
the very weak nature of these electronic modulations and their strong 
fluctuations\cite{Wise2008}.

In recent years there was an enormous improvement in the precision of 
the CO wavelength $\lambda_{\rm CO}$ measurements specially by STM, x-ray 
and REXS\cite{Comin2016}.
The very fine variation of $\lambda_{\rm CO}$ on $p$ revealed in these
experiments suggests the use of the time-dependent nonlinear Cahn-Hilliard (CH)
differential equation\cite{Cahn1958}. In this approach
the different charge oscillations may be tuned slowly
up to reproduce the measured $\lambda_{\rm CO}$
and other forms of
alternating hole-rich and hole poor regions on 100\% volume
fraction\cite{deMello2009,DeMello2012,deMelloKasal2012,DeMello2014,Mello2017}.

The method
has also the great advantage to concomitantly provide the free energy
that yields a connection between the SC interaction and the charge modulations. 
The starting point is the time-dependent
phase separation order parameter associated with the local electronic
density, $u({\bf r},t) = (p({\bf r},t) - p)/p$, where 
$p({\bf r},t)$ is the  local charge or hole density at a 
position ${\bf r}$ in the plane. The CH equation is based on the
Ginzburg-Landau (GL) free energy expansion in terms of the conserved order parameter $u$    \cite{Otton2005,deMello2009,DeMello2012,deMelloKasal2012,DeMello2014,Mello2017}: 
\begin{equation}
f(u)= {{\frac{1}{2}\varepsilon |\nabla u|^2 +V_{\rm GL}(u,T)}},
\label{FE}
\end{equation}
where $\varepsilon$ is the parameter that controls the charge modulations
and ${V_{\rm GL}}(u,T)= -\alpha [T_{\rm PS}-T] u^2/2+B^2u^4/4+...$ is a
double-well potential that characterizes the rise of charge oscillations 
below $T_{\rm PS}$ that is near $T^*(p)$. In general the values of 
$\alpha$ and $B$ are equated to unity. This free energy in terms of the phase 
separation order
parameter is much simpler than the Ginzburg-Landau-Wilson free energy in 
terms of SC and PDW fields\cite{InterOrder2015}  but it suitably reproduces
the details of the CO structure of distinct compounds and their localization
energy ${V_{\rm GL}}$.

The CH equation can be written in the form of the following continuity equation 
for the local free energy current density  ${\bf J}=M{\bf\nabla}(\delta f/ \delta u)$,\cite{Bray1994}
\begin{eqnarray}
\frac{\partial u}{\partial t} & = & -{\bf \nabla.J} \nonumber \\
& = & -M\nabla^2[\varepsilon^2\nabla^2u
- \alpha^2(T)u+B^2u^3],
\label{CH}
\end{eqnarray}
where $M$ is the mobility or the charge transport coefficient that sets both the phase separation time scale and
the contrast between the values of $u$ for the two phases. 

The equation is solved by a stable and 
fast finite difference scheme  with free boundary
conditions\cite{Otton2005} yielding the 
phase separation conserved order parameter $u({\bf r},t = n\delta t)$,
function of position ${\bf r}$ and $n$ simulation time step $\delta t$.
The limiting cases are
$u({\bf r}_i,t)\approx 0$ corresponding to homogeneous systems above 
the pseudogap temperature $T^*$
or small charge variations like the observed charge density wave (CDW) and
$u({\bf r}_i,t\rightarrow \infty) =  \pm  \:1$, corresponding to the extreme case 
(at low temperatures) of complete phase separation. 
The local charge density is derived from
$p({\bf r},t) = p \times (u({\bf r},t) + 1)$ and the later case (strong phase separation) 
applies to static 
stripes\cite{Tranquada1995a,Thampy2017} while the former (weak phase separation) to very small 
$\Delta p \approx 10^{-2-3}$ variations around $p$, like that measured 
in YBa$_2$Cu$_3$O$_{6+\delta}$ (Y123)\cite{Kharkov2016}.
We believe that such weak charge modulation masked CDW for many
years and it is probably the reason to the very few charge inhomogeneities 
observations in the overdoped regime\cite{Wu2017,CDWover2018}.

\begin{figure}
\includegraphics[height=7.50cm]{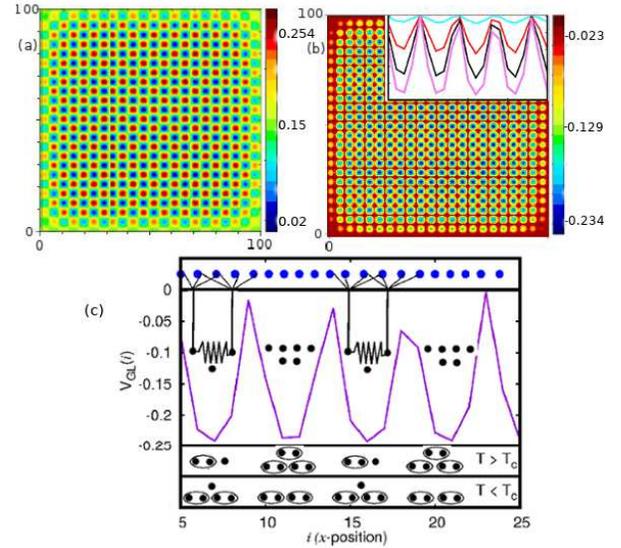}
\caption{
(a) Low temperature simulation of a checkerboard charge density
pattern for LSCO $p = 0.15$ on 100 vs.100 sites and (b) the $V_{\rm GL}(u({\bf r}))$ 
that yields this density map. The inset shows how $V_{\rm GL}(x)$ evolves 
with the temperature below $T^*$ or with 
the time of simulation $\delta t$. (c) The $T = 0$ K limit of $V_{\rm GL}(x)$.
At the  
top, we represent some planar Cu atoms (blue filled circles) attracted to the hole-rich
regions represented by straight lines as an illustration. Hole motion in the domains
produces atomic fluctuations that affect other holes promoting an atomic 
mediated interaction that is represented by the 
springs for illustration. At $T \le T_{\rm c}$ long-range order sets in,
the Cooper pairs (the encircled pair of black dots) spread (superflowing) and the CO x-ray
scattering signal decreases\cite{Chang2012,Wise2008}.
}
\label{fig1}
\end{figure}

Figure \ref{fig1}(a)
shows a typical CH simulation (La$_{\rm 2-x}$Sr$_{\rm x}$CuO$_4$
(LSCO) for $p = 0.15$)
with a checkerboard pattern of $\lambda_{\rm CO} \sim 4 a_0$ where $a_0$
is the lattice  parameter, while
Fig. \ref{fig1}(b) shows $V_{\rm GL}(u({\bf r}))$ or just 
$V_{\rm GL}({\bf r})$ map that
originates this specific charge structure.
We discussed already these simulations in detail\cite{DeMello2014,Mello2017} 
and {\it here we want to focus mainly on the SC interaction} promoted by
 $V_{\rm GL}({\bf r})$ that has a double role: First,
it generates non-uniform charge patterns like the 
checkerboard modulation displayed in Fig. \ref{fig1}(a) that affect the 
ionic electronic clouds. The small atomic oscillatory displacements 
are verified by several neutron and x-ray scattering experiments\cite{Abbamonte2006,Chang2012}. 
Second, the small amplitude 
rapidly varying $\Delta p$ are transmitted to the 
atomic electronic clouds that transmit them back to the holes, generating 
a hole-hole lattice mediated interaction.

In Fig. \ref{fig1}(c) we plot $V_{\rm GL}(x)$ along the $x$-direction 
together with some localized holes represented by the black filled circles. 
The planar Cu atoms are represented schematically on the top of Fig. \ref{fig1}(c)
by the filled blue circles slightly displaced to (from) the hole-rich(poor) domains.
When the temperature decreases below $T^*$ the $V_{\rm GL}({\bf r})$
modulations increase as shown schematically in the inset of Fig. \ref{fig1}(b)
favoring alternating charge  domains like those of Fig.\ref{fig1}(a). 
These domains are large compared with $a_0$ and if the $V_{\rm GL}({\bf r})$ modulations
are high enough, the holes may move around or oscillate in the domains
what induce also 
fluctuations on the nearby Cu atoms that,
like a mirror, interact back with the other holes in the same domain.  

This process leads to our main assumption; the SC local hole-hole pairing 
interaction is proportional to the  spatial average
${\left < V_{\rm GL}(p,0) \right >} \equiv \sum_i^N V_{\rm GL}(r_i,p,0)/N$, where
the sum is over all the planar sites. The indirect role of the lattice 
in the SC interaction is confirmed by the
relatively large isotope effect on the onset of superconductivity below $T^*$ 
and also on $T_{\rm c}$\cite{Isotop2017}.

With this phenomenological potential we developed a 
particular type of Bogoliubov-deGennes (BdG) SC approach that 
converges self-consistently to the local chemical potential $\mu_i$ and the local
$d$-wave amplitude $\Delta_d( r_i)$, keeping always the original CO structure
fixed\cite{deMello2009,deMelloKasal2012,DeMello2012,DeMello2014,Mello2017}.
This is done diagonalizing
the BdG matrix with the Hubbard Hamiltonian with hopping parameters taken from 
ARPES and nearest neighbor attractive potential with
temperature dependence from the GL method; 
$V_{\rm GL}(p, T)={\left < V_{\rm GL}(p, 0)\right >}[1-T/T^*]^2$.

Notice that $V_{\rm GL}(p, T)$ is defined as a function of the
dimensionless phase separation order parameter $u({\bf r},t)$ (Eq. \ref{FE}) and needs to
be multiplied by a dimensional constant to be converted 
to energy units in the Hubbard Hamiltonian. 
This parameter is obtained making the low temperature CH-BdG calculations with the 
attractive potential ${\left < V_{\rm GL}(p, 0) \right >}$  
in meV to yield 
the experimental optimal gap $\left < \Delta_d(p_{\rm opt} = 0.16, 0)\right >$
also in meV. 
The very same constant is used to the other compounds potential
${\left < V_{\rm GL}(p, 0) \right >}$, what gives only one adjustable
parameter to all $\left < \Delta_d(p ,T)\right >$ of a given family.


These CH-BdG calculations on a charge density map like that of
Fig. \ref{fig1}(a) yield local SC amplitudes $\Delta_d(r_i)$ 
inside each charge domain, in agreement
with typical SC coherence length $\xi_{\rm SC}$\cite{Lang2002,Imry2012,DaSilvaNeto2014}  
smaller\cite{ECarlson2002} 
than typical $\lambda_{\rm CO}$\cite{Comin2016}.
The $\Delta_d(r_i)$ plots have
the same modulations of the charges, what is known as PDW, but in 
our approach this is a natural consequence of the simultaneous self-consistent
approach on $\mu_i$ and $\Delta_d( r_i)$\cite{Mello2017}.  
This is a different view of the proposal that PDW is a ``parent'' phase which
spontaneously break  symmetries and gives rise to the CDW and SC orders\cite{InterOrder2015}.

To extend this approach to the overdoped region we recall our pillar
connecting the CO with the PG and that $T^*(p)$ vanishes only at
$p \approx 0.27$, the end of the SC dome\cite{Ando2004,InterOrder2015,Huefner2007}. 
This argument suggests that weak incommensurate
charge modulations are also present, most likely with much weaker amplitudes, 
in the overdoped region. In fact,
different types of inhomogeneities are observed in overdoped Bi-based 
families\cite{STM.Do2008,Gomes2007} and in La-based materials\cite{Yoshida2012}
that is possibly connected with charge instabilities.
Electronic transport anisotropy in the CuO plane that decreases with temperature
and doping, persisting up to at least $p \sim 0.22$ was recently measured and 
associated with a nematic phase\cite{Wu2017}.
More recently, REXS experiments  
in strong overdoped Bi2201 observed CO peak signals similar to those of
underdoped cuprates\cite{CDWover2018}. They also measured a continued decrease
of the CO vector $Q_{\rm CO}$ versus doping similar to what is seen in underdoped 
compounds\cite{Comin2016}.

\begin{figure}
\centerline{ \includegraphics[height=4.50cm]{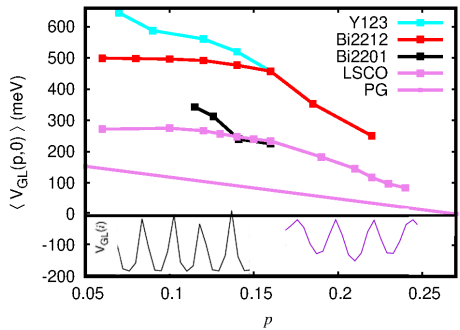}}
\caption{
 The SC pair potential  ${\left < V_{\rm GL} (p,0)\right >}$ derived from the 
CO maps like
that of Fig.\ref{fig1}(b) converted in energy units (meV)
to reproduce (with the BdG calculations) the  
SC gaps $\Delta_0(p)$ in agreement with the experiments. We plot also the $\Delta_{\rm PG}(p)$  
from Ref. [\onlinecite{Huefner2007}] to llustrate its similar dependence on $p$ with those of the averages ${\left < V_{\rm GL} (p,0)\right >}$. In the bottom, we
show schematically the $V_{\rm GL}$ amplitude characteristic  of the underdoped, 
respectively of the overdoped regions that are correlated with the SC interaction.
}
\label{VGLDd}
\end{figure}

\begin{figure}
\centerline{ \includegraphics[height=5.0cm]{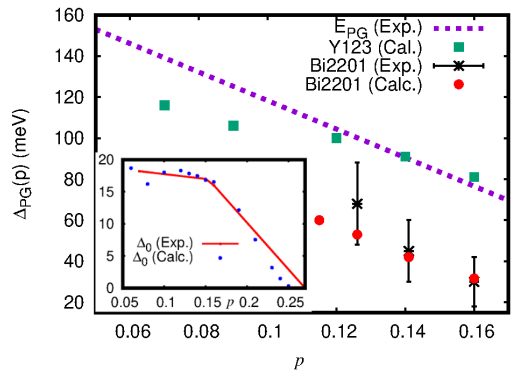}}
\caption{ The 
$\Delta_{\rm PG}$ of five Y123 and four Bi2201 calculated by Eq. \ref{PG} 
with  experimental values 
of $\lambda_{\rm CO}$ compiled by Ref. [\onlinecite{Comin2016}] and  
the ${\left < V_{\rm GL} (p,0)\right >}$ potential values from Fig. \ref{VGLDd}, without any 
adjusted parameter. The dashed line is an average of the experimental data 
from Ref. [\onlinecite{Huefner2007}] and 
the Bi2201 experimental points and errorbars are from Ref. [\onlinecite{Wise2008}].
In the inset, the SC energy scale, 
${\left < \Delta_0 (p, 0)\right >}$ from the BdG calculations with
the ${\left < V_{\rm GL} (p,0)\right >}$ potential of LSCO.
}
\label{PGgaps}
\end{figure}

Taking these observations into consideration and the
$\lambda_{\rm CO}(p)$ data of La and Bi-based compounds
we extend the calculation to the overdoped region.
The values of ${\left < V_{\rm GL} (p, 0)\right >}$ are plotted in Fig. \ref{VGLDd} 
and used again in the BdG calculations to derive the low temperature SC gaps 
$\left < \Delta_d(p, 0)\right >$ or $\Delta_0(p)$ plotted
in the inset of Fig. \ref{PGgaps} for both under and overdoped
La$_{\rm 2-x}$Sr$_{\rm x}$CuO$_4$ (LSCO). 
The results are close to the measured ARPES nodal gaps $\Delta_0(p)$ extrapolated
to the antinodal direction\cite{Lee2007, Yoshida2012}, specific heat\cite{Tallon2003} 
and STM\cite{STM.Do2008} measurements, indicating that $\Delta_0(p)$ is a good
candidate to the SC gap $\Delta_{\rm SC}(p)$.

Now, we have the ingredients to demonstrate the correlation between 
the PG and the CO through the derivation
of $\Delta_{\rm PG}(p)$ for Y123 and Bi2201 using the measured
$\lambda_{\rm CO}(p)$ compiled in Ref.[\onlinecite{Comin2016}] and 
the ${\left < V_{\rm GL} (p,0)\right >}$ that we derived and plotted in Fig. \ref{VGLDd}.
We equate $\Delta_{\rm PG}(p,0)$ to the ground state energy of a 
shallow 2D well
$U ={\left < V_{\rm GL} (p,0)\right >}$\cite{LMQ}:

\begin{equation}
 \Delta_{\rm PG} = \frac{\hbar^2}{m\lambda_{\rm CO}^2}exp 
 [-\frac{2\hbar^2}{m\lambda_{\rm CO}^2U}] .
 \label{PG}
\end{equation}
The results of five Y123 and four Bi2201 calculations
are plotted in Fig. \ref{PGgaps} together with the experimental data\cite{Huefner2007, Wise2008}. 
The agreement near optimal doping
is almost perfect and we emphasize that there is not any adjusted parameter in Eq. \ref{PG}.

\begin{figure}
 \includegraphics[height=6.5cm]{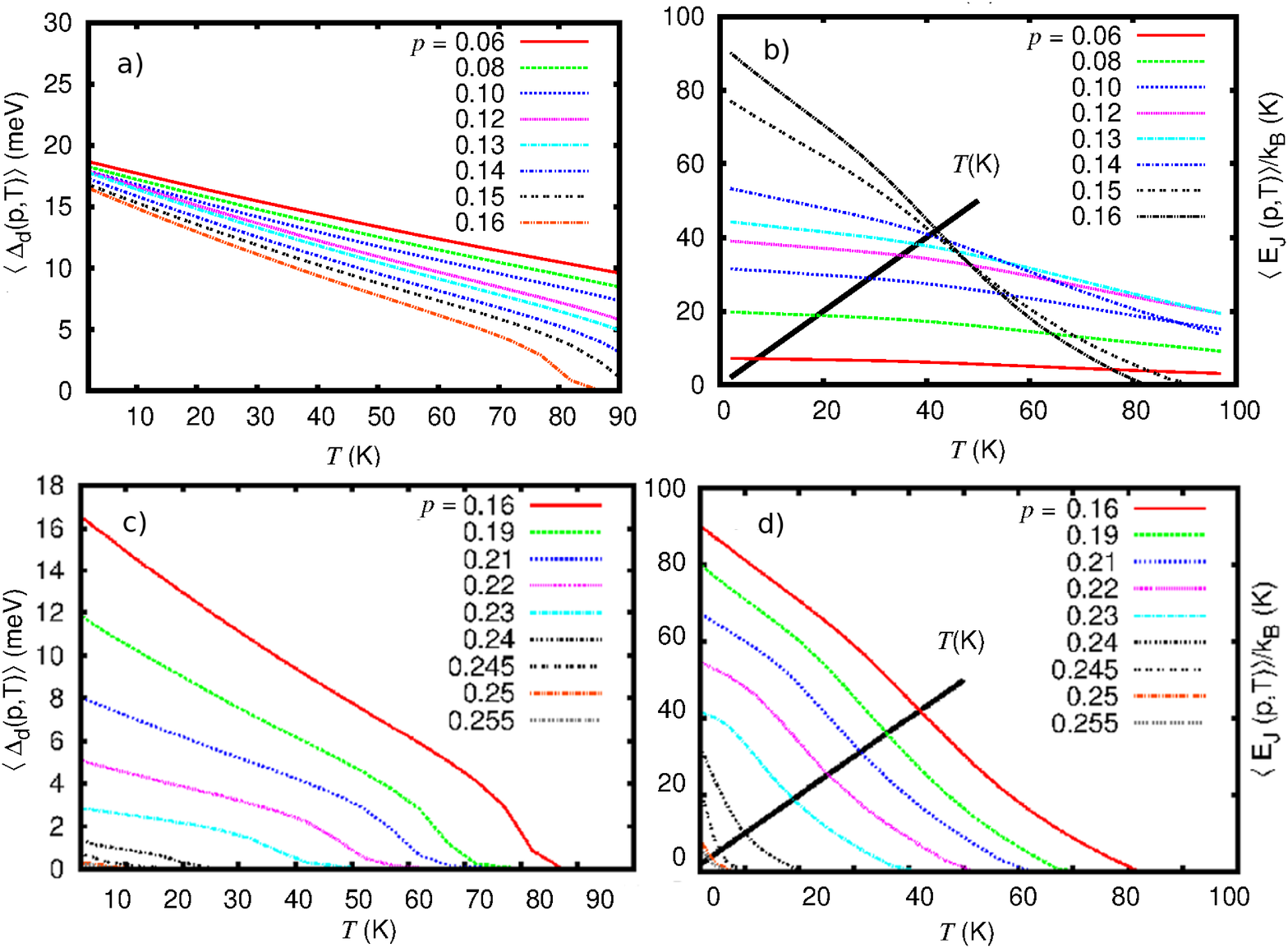}
\caption{a) ${\left < \Delta_d(p,T)\right >}$ for underdoped LSCO 
starting with the approximately constant $\left < \Delta_0(p,0)\right >\approx17.5$meV 
that are close to the measured maximum gap $\Delta_0(p)$ as shown in the inset
of Fig. \ref{PGgaps}. 
b) Josephson energy ${\left < E_{\rm J}(p,T) \right >}$ derived from the 
gaps shown in a). The 
intersections with $k_{\rm B}T$ yields the two sets of $T_{\rm c}(p)$.
c) The same of a) for overdoped LSCO compounds. d) Josephson energy 
${\left < E_{\rm J}(p,T) \right >}$ derived from the 
gaps shown in c). 
}
\label{figD1T}
\end{figure}

The agreement of the PG calculations endorses the CH-BdG calculations of 
CDW like structures with local SC order parameters inside
the charge domains. In this scenario, the SC properties are similar
to those of granular superconductors with Cooper pairs tunneling\cite{Mello2017}. 
Such model was proposed earlier to explain the distribution of localized 
gaps detected by STM\cite{Lang2002} and the SC correlations above 
$T_{\rm c}(p)$ in several materials\cite{Imry2012}.
In general, a SC order parameter has two components, ($\Delta_d(r_i)$, $\theta_i$) 
what leads to the superconductivity in two steps and 
provides an explanations to the SC correlations measured at temperatures above $T_{\rm c}$\cite{Gomes2007,Kanigel2008,Dubroka2011,Bilbro2011}. 
These experiments and our calculations of finite $\Delta_d(r_i)$ and finite
${\left < \Delta_0 (p, T)\right >}$ above $T_{\rm c}$ (shown,
for instance, in Fig. \ref{figD1T}(a) and (c))
suggest that the system resistance just above the SC transition 
comes from the persistent normal regions
between the SC domains and their boundaries that oppose the Cooper pairs
tunneling\cite{Mello2017}. 

In this case $T_{\rm c}(p)$ is the long-range phase order temperature
obtained by Josephson coupling between the phases $\theta_i$ in the 
charge domains.
We have explained previously\cite{DeMello2014} that for a $d$-wave superconductor
junction is sufficient to use the following $s$-wave analytical
average Josephson coupling expression
\begin{equation}
 {\left < E_{\rm J}(p,T) \right >} = \frac{\pi \hbar {\left <\Delta_d(p,T)\right >}}
 {2 e^2 R_{\rm n}(p)} 
 {\rm tanh} \bigl [\frac{\left <\Delta_d(p,T)\right >}{2k_{\rm B}T} \bigr ] ,
\label{EJ} 
\end{equation}
where $R_{\rm n}(p)$ is taken to be proportional to the 
$T \gtrapprox T_{\rm c}$ normal-state in-plane resistivity 
$\rho_{ab}(p)$ obtained from typical $\rho_{ab}(p,T)$ vs. $T$ curves\cite{Ando2004}. 
The proportionality constant between $R_{\rm n}$ and $\rho_{ab}$ is found matching
the optimal  $T_{\rm c} \approx 42$ K for the case of LSCO. 
The other LSCO compounds use the same
constant so that we need just a single adjustable parameter to derive
$T_{\rm c}(p)$ for each cuprate system.

\begin{figure}
\centerline{ \includegraphics[height=5.0cm]{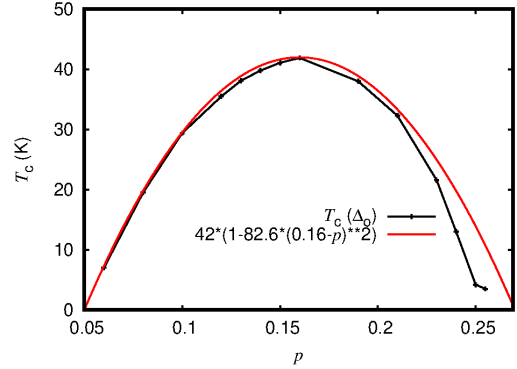}}
\caption{$T_{\rm c} (p)$ calculation on the whole 
hole-doping region from the $E_{\rm J}$ of Eq. \ref{EJ} and
Figs. \ref{figD1T}(b) (underdoped) and (d) (overdoped).
The results reproduce the well-known LSCO measurements.
}
\label{3gaps}
\end{figure}

In Fig. \ref{figD1T}(b) and (d) we plot ${\left < E_{\rm J}(p,T) \right >}$ whose
intersections with $k_{\rm B} T$, represented by the black 
straight lines, yield $T_{\rm c}(p)$. The results shown in Fig. \ref{3gaps} 
comprise all the CH-BdG
calculations described previously in this paper yielding
under and overdoped $T_{\rm c}(p)$ for the LSCO case. 
The agreement with the
experiments is almost perfect, it deviates only in the strong overdoped region
where the PG 
vanishes, the system become almost uniform and the ${\left < E_{\rm J}(p,T) \right >}$ 
uncertainty is large. The $T_{\rm c}(p)$ dome shape has a simple interpretation with
Eq. \ref{EJ}; the
competing contribution of ${\left <\Delta_d(p,0)\right >}$ 
that decreases steadily with $p$ and vanishes near $p = 0.27$ together with 
$T^*(p)$ (see inset of Fig. \ref{PGgaps}), and $R_{\rm n}(p)$ 
that has an exponential decreasing behavior and diverges near $p = 0.05$. 
 We should mention that other methods were also successful to
reproduce the $T_{\rm c}(p)$ bell shape of cuprates, in particular, 
the method of Green function 
techniques with rigorous implementation of the non-standard commutation 
relations of the Hubbard operators\cite{Plakida2003,Plakida2016}.

Another novel interpretation that comes out 
of this approach is that the Cooper pairs
acquire long-range order at $T \sim T_{\rm c}$ and like a superfluid
spreads over the charge modulation domains on 
the CuO planes. This superflow uniforms the total
charge density leading to a substantial decrease of the CO
interference x-ray scattering signal. {\it This  was
interpreted as due to the competition between the CO and the SC phase\cite{Chang2012,Comin2015a},}
but it is a consequence of the local Cooper pairs long-range SC transition, 
as it is schematically shown at the bottom of Fig. \ref{fig1}(c) 
(for $T>T_{\rm c}$ and $T<T_{\rm c}$).

We have shown how to calculate the pseudogap $\Delta_{\rm PG}(p)$, the average SC gap 
${\left < \Delta_d (p) \right >}$ and $T_{\rm c} (p)$ of different cuprates
in very good agreement with experiments from the measurements of $\lambda_{\rm CO}(p)$. 
 We remark that  Ref. [\onlinecite{InterOrder2015}] proposes that the 
PDW order promotes other symmetry breaking phases of cuprates, in this context, 
we can say that our calculations show that the CDW order promotes 
the PG and SC $d$-wave phases.
To deal with many important
properties discussed in the paper and reproduce accurately several quantitative results,
our approach is essentially phenomenological but should provide clear guidelines 
to any fundamental theoretical calculation on cuprates.
Another advantage of our phenomenological theory is to reveal in a simple way the connection
between the most fundamental energy scales and the relation between distinct properties
like, for instance, the PG and the SC interaction. 

The method is general and simple to be used in any problem involving 
superconductivity with CDW or any
other type of charge instability that, otherwise do not have a simple theoretical approach.  
Under this program, we will soon present calculations on
the correlation between the superfluid density $\rho_{\rm sc}(p,0)$ 
and $T_{\rm c}(p)$\cite{Mello2020b}, the interpretation of high magnetic field
quantum oscillations experiments\cite{Mello2020c}, proximity effects, 
and other challenging problems 
of cuprates. Eq. \ref{EJ} also points the way to combine materials to produce 
larger values of $T_{\rm c}$ what is important to technological applications.

I thank A. Bianconi, I. Bo{\v{z}}ovi{\'c}, D. M\"ockli, J. Tranquada for helpful 
discussions and partial support by the Brazilian agencies CNPq and FAPERJ.


\begin{thebibliography}{59}%
\makeatletter
\providecommand \@ifxundefined [1]{%
 \@ifx{#1\undefined}
}%
\providecommand \@ifnum [1]{%
 \ifnum #1\expandafter \@firstoftwo
 \else \expandafter \@secondoftwo
 \fi
}%
\providecommand \@ifx [1]{%
 \ifx #1\expandafter \@firstoftwo
 \else \expandafter \@secondoftwo
 \fi
}%
\providecommand \natexlab [1]{#1}%
\providecommand \enquote  [1]{``#1''}%
\providecommand \bibnamefont  [1]{#1}%
\providecommand \bibfnamefont [1]{#1}%
\providecommand \citenamefont [1]{#1}%
\providecommand \href@noop [0]{\@secondoftwo}%
\providecommand \href [0]{\begingroup \@sanitize@url \@href}%
\providecommand \@href[1]{\@@startlink{#1}\@@href}%
\providecommand \@@href[1]{\endgroup#1\@@endlink}%
\providecommand \@sanitize@url [0]{\catcode `\\12\catcode `\$12\catcode
  `\&12\catcode `\#12\catcode `\^12\catcode `\_12\catcode `\%12\relax}%
\providecommand \@@startlink[1]{}%
\providecommand \@@endlink[0]{}%
\providecommand \url  [0]{\begingroup\@sanitize@url \@url }%
\providecommand \@url [1]{\endgroup\@href {#1}{\urlprefix }}%
\providecommand \urlprefix  [0]{URL }%
\providecommand \Eprint [0]{\href }%
\providecommand \doibase [0]{http://dx.doi.org/}%
\providecommand \selectlanguage [0]{\@gobble}%
\providecommand \bibinfo  [0]{\@secondoftwo}%
\providecommand \bibfield  [0]{\@secondoftwo}%
\providecommand \translation [1]{[#1]}%
\providecommand \BibitemOpen [0]{}%
\providecommand \bibitemStop [0]{}%
\providecommand \bibitemNoStop [0]{.\EOS\space}%
\providecommand \EOS [0]{\spacefactor3000\relax}%
\providecommand \BibitemShut  [1]{\csname bibitem#1\endcsname}%
\let\auto@bib@innerbib\@empty
\bibitem [{\citenamefont {Fradkin}\ \emph {et~al.}(2015)\citenamefont
  {Fradkin}, \citenamefont {Kivelson},\ and\ \citenamefont
  {Tranquada}}]{InterOrder2015}%
  \BibitemOpen
  \bibfield  {author} {\bibinfo {author} {\bibfnamefont {E.}~\bibnamefont
  {Fradkin}}, \bibinfo {author} {\bibfnamefont {S.~A.}\ \bibnamefont
  {Kivelson}}, \ and\ \bibinfo {author} {\bibfnamefont {J.~M.}\ \bibnamefont
  {Tranquada}},\ }\href {\doibase 10.1103/RevModPhys.87.457} {\bibfield
  {journal} {\bibinfo  {journal} {Rev. Mod. Phys.}\ }\textbf {\bibinfo {volume}
  {87}},\ \bibinfo {pages} {457} (\bibinfo {year} {2015})}\BibitemShut
  {NoStop}%
\bibitem [{\citenamefont {Huefner}\ \emph {et~al.}(2008)\citenamefont
  {Huefner}, \citenamefont {Hossain}, \citenamefont {Damascelli},\ and\
  \citenamefont {Sawatzky}}]{Huefner2007}%
  \BibitemOpen
  \bibfield  {author} {\bibinfo {author} {\bibfnamefont {S.}~\bibnamefont
  {Huefner}}, \bibinfo {author} {\bibfnamefont {M.~A.}\ \bibnamefont
  {Hossain}}, \bibinfo {author} {\bibfnamefont {A.}~\bibnamefont {Damascelli}},
  \ and\ \bibinfo {author} {\bibfnamefont {G.~A.}\ \bibnamefont {Sawatzky}},\
  }\href {http://arxiv.org/abs/0706.4282} {\bibfield  {journal} {\bibinfo
  {journal} {Rep. Prog. Phys.}\ }\textbf {\bibinfo {volume} {71}},\ \bibinfo
  {pages} {062501} (\bibinfo {year} {2008})}\BibitemShut {NoStop}%
\bibitem [{\citenamefont {Munnikes}\ \emph {et~al.}(2011)\citenamefont
  {Munnikes} \emph {et~al.}}]{Raman2011}%
  \BibitemOpen
  \bibfield  {author} {\bibinfo {author} {\bibfnamefont {N.}~\bibnamefont
  {Munnikes}} \emph {et~al.},\ }\href {\doibase 10.1103/PhysRevB.84.144523}
  {\bibfield  {journal} {\bibinfo  {journal} {Phys. Rev. B}\ }\textbf {\bibinfo
  {volume} {84}},\ \bibinfo {pages} {144523} (\bibinfo {year}
  {2011})}\BibitemShut {NoStop}%
\bibitem [{\citenamefont {Tallon}\ \emph {et~al.}(2003)\citenamefont {Tallon},
  \citenamefont {Loram}, \citenamefont {Cooper}, \citenamefont {Panagopoulos},\
  and\ \citenamefont {Bernhard}}]{Tallon2003}%
  \BibitemOpen
  \bibfield  {author} {\bibinfo {author} {\bibfnamefont {J.~L.}\ \bibnamefont
  {Tallon}}, \bibinfo {author} {\bibfnamefont {J.~W.}\ \bibnamefont {Loram}},
  \bibinfo {author} {\bibfnamefont {J.~R.}\ \bibnamefont {Cooper}}, \bibinfo
  {author} {\bibfnamefont {C.}~\bibnamefont {Panagopoulos}}, \ and\ \bibinfo
  {author} {\bibfnamefont {C.}~\bibnamefont {Bernhard}},\ }\href {\doibase
  10.1103/PhysRevB.68.180501} {\bibfield  {journal} {\bibinfo  {journal} {Phys.
  Rev. B}\ }\textbf {\bibinfo {volume} {68}},\ \bibinfo {pages} {180501}
  (\bibinfo {year} {2003})}\BibitemShut {NoStop}%
\bibitem [{\citenamefont {Yoshida}\ \emph {et~al.}(2012)\citenamefont
  {Yoshida}, \citenamefont {Hashimoto}, \citenamefont {Vishik}, \citenamefont
  {Shen},\ and\ \citenamefont {Fujimori}}]{Yoshida2012}%
  \BibitemOpen
  \bibfield  {author} {\bibinfo {author} {\bibfnamefont {T.}~\bibnamefont
  {Yoshida}}, \bibinfo {author} {\bibfnamefont {M.}~\bibnamefont {Hashimoto}},
  \bibinfo {author} {\bibfnamefont {I.~M.}\ \bibnamefont {Vishik}}, \bibinfo
  {author} {\bibfnamefont {Z.-X.}\ \bibnamefont {Shen}}, \ and\ \bibinfo
  {author} {\bibfnamefont {A.}~\bibnamefont {Fujimori}},\ }\href {\doibase
  10.1143/JPSJ.81.011006} {\bibfield  {journal} {\bibinfo  {journal} {J. Phys.
  Soc. Japan}\ }\textbf {\bibinfo {volume} {81}},\ \bibinfo {pages} {011006}
  (\bibinfo {year} {2012})}\BibitemShut {NoStop}%
\bibitem [{\citenamefont {Hashimoto}\ \emph {et~al.}(2007)\citenamefont
  {Hashimoto} \emph {et~al.}}]{Yoshida2007}%
  \BibitemOpen
  \bibfield  {author} {\bibinfo {author} {\bibfnamefont {M.}~\bibnamefont
  {Hashimoto}} \emph {et~al.},\ }\href {\doibase 10.1103/PhysRevB.75.140503}
  {\bibfield  {journal} {\bibinfo  {journal} {Phys. Rev. B}\ }\textbf {\bibinfo
  {volume} {75}},\ \bibinfo {pages} {140503} (\bibinfo {year}
  {2007})}\BibitemShut {NoStop}%
\bibitem [{\citenamefont {Kato}\ \emph {et~al.}(2008)\citenamefont {Kato},
  \citenamefont {Maruyama}, \citenamefont {Okitsu},\ and\ \citenamefont
  {Sakata}}]{Kato2008}%
  \BibitemOpen
  \bibfield  {author} {\bibinfo {author} {\bibfnamefont {T.}~\bibnamefont
  {Kato}}, \bibinfo {author} {\bibfnamefont {T.}~\bibnamefont {Maruyama}},
  \bibinfo {author} {\bibfnamefont {S.}~\bibnamefont {Okitsu}}, \ and\ \bibinfo
  {author} {\bibfnamefont {H.}~\bibnamefont {Sakata}},\ }\href {\doibase
  10.1143/JPSJ.77.054710} {\bibfield  {journal} {\bibinfo  {journal} {Journal
  of the Physical Society of Japan}\ }\textbf {\bibinfo {volume} {77}},\
  \bibinfo {pages} {054710} (\bibinfo {year} {2008})}\BibitemShut {NoStop}%
\bibitem [{\citenamefont {Alldredge}\ \emph {et~al.}(2008)\citenamefont
  {Alldredge} \emph {et~al.}}]{STM.Do2008}%
  \BibitemOpen
  \bibfield  {author} {\bibinfo {author} {\bibfnamefont {J.~W.}\ \bibnamefont
  {Alldredge}} \emph {et~al.},\ }\href {https://doi.org/10.1038/nphys917}
  {\bibfield  {journal} {\bibinfo  {journal} {Nature Physics}\ }\textbf
  {\bibinfo {volume} {4}},\ \bibinfo {pages} {319 EP } (\bibinfo {year}
  {2008})}\BibitemShut {NoStop}%
\bibitem [{\citenamefont {Ren}\ \emph {et~al.}(2012)\citenamefont {Ren} \emph
  {et~al.}}]{Tunnel2012}%
  \BibitemOpen
  \bibfield  {author} {\bibinfo {author} {\bibfnamefont {J.~K.}\ \bibnamefont
  {Ren}} \emph {et~al.},\ }\href {https://doi.org/10.1038/srep00248} {\bibfield
   {journal} {\bibinfo  {journal} {Scientific Reports}\ }\textbf {\bibinfo
  {volume} {2}},\ \bibinfo {pages} {248 EP } (\bibinfo {year}
  {2012})}\BibitemShut {NoStop}%
\bibitem [{\citenamefont {Comin}\ and\ \citenamefont
  {Damascelli}(2016)}]{Comin2016}%
  \BibitemOpen
  \bibfield  {author} {\bibinfo {author} {\bibfnamefont {R.}~\bibnamefont
  {Comin}}\ and\ \bibinfo {author} {\bibfnamefont {A.}~\bibnamefont
  {Damascelli}},\ }\href {\doibase 10.1146/annurev-conmatphys-031115-011401}
  {\bibfield  {journal} {\bibinfo  {journal} {Ann. Rev. of Cond. Mat. Phys.}\
  }\textbf {\bibinfo {volume} {7}},\ \bibinfo {pages} {369} (\bibinfo {year}
  {2016})}\BibitemShut {NoStop}%
\bibitem [{\citenamefont {Wise}\ \emph {et~al.}(2008)\citenamefont {Wise} \emph
  {et~al.}}]{Wise2008}%
  \BibitemOpen
  \bibfield  {author} {\bibinfo {author} {\bibfnamefont {W.~D.}\ \bibnamefont
  {Wise}} \emph {et~al.},\ }\href {\doibase 10.1038/nphys1021} {\bibfield
  {journal} {\bibinfo  {journal} {Nature Physics}\ }\textbf {\bibinfo {volume}
  {4}},\ \bibinfo {pages} {696} (\bibinfo {year} {2008})}\BibitemShut {NoStop}%
\bibitem [{\citenamefont {Comin}\ \emph {et~al.}(2014)\citenamefont {Comin}
  \emph {et~al.}}]{Comin2014}%
  \BibitemOpen
  \bibfield  {author} {\bibinfo {author} {\bibfnamefont {R.}~\bibnamefont
  {Comin}} \emph {et~al.},\ }\href {\doibase 10.1126/science.1242996}
  {\bibfield  {journal} {\bibinfo  {journal} {Science (New York, N.Y.)}\
  }\textbf {\bibinfo {volume} {343}},\ \bibinfo {pages} {390} (\bibinfo {year}
  {2014})}\BibitemShut {NoStop}%
\bibitem [{\citenamefont {Xia}\ \emph {et~al.}(2008)\citenamefont {Xia} \emph
  {et~al.}}]{Kerr2008}%
  \BibitemOpen
  \bibfield  {author} {\bibinfo {author} {\bibfnamefont {J.}~\bibnamefont
  {Xia}} \emph {et~al.},\ }\href {\doibase 10.1103/PhysRevLett.100.127002}
  {\bibfield  {journal} {\bibinfo  {journal} {Phys. Rev. Lett.}\ }\textbf
  {\bibinfo {volume} {100}},\ \bibinfo {pages} {127002} (\bibinfo {year}
  {2008})}\BibitemShut {NoStop}%
\bibitem [{\citenamefont {Lubashevsky}\ \emph {et~al.}(2014)\citenamefont
  {Lubashevsky}, \citenamefont {Pan}, \citenamefont {Kirzhner}, \citenamefont
  {Koren},\ and\ \citenamefont {Armitage}}]{OptBirefri2014}%
  \BibitemOpen
  \bibfield  {author} {\bibinfo {author} {\bibfnamefont {Y.}~\bibnamefont
  {Lubashevsky}}, \bibinfo {author} {\bibfnamefont {L.}~\bibnamefont {Pan}},
  \bibinfo {author} {\bibfnamefont {T.}~\bibnamefont {Kirzhner}}, \bibinfo
  {author} {\bibfnamefont {G.}~\bibnamefont {Koren}}, \ and\ \bibinfo {author}
  {\bibfnamefont {N.~P.}\ \bibnamefont {Armitage}},\ }\href {\doibase
  10.1103/PhysRevLett.112.147001} {\bibfield  {journal} {\bibinfo  {journal}
  {Phys. Rev. Lett.}\ }\textbf {\bibinfo {volume} {112}},\ \bibinfo {pages}
  {147001} (\bibinfo {year} {2014})}\BibitemShut {NoStop}%
\bibitem [{\citenamefont {Mahyari}\ \emph {et~al.}(2013)\citenamefont {Mahyari}
  \emph {et~al.}}]{Muon2013}%
  \BibitemOpen
  \bibfield  {author} {\bibinfo {author} {\bibfnamefont {Z.~L.}\ \bibnamefont
  {Mahyari}} \emph {et~al.},\ }\href {\doibase 10.1103/PhysRevB.88.144504}
  {\bibfield  {journal} {\bibinfo  {journal} {Phys. Rev. B}\ }\textbf {\bibinfo
  {volume} {88}},\ \bibinfo {pages} {144504} (\bibinfo {year}
  {2013})}\BibitemShut {NoStop}%
\bibitem [{\citenamefont {Gomes}\ \emph {et~al.}(2007)\citenamefont {Gomes}
  \emph {et~al.}}]{Gomes2007}%
  \BibitemOpen
  \bibfield  {author} {\bibinfo {author} {\bibfnamefont {K.~K.}\ \bibnamefont
  {Gomes}} \emph {et~al.},\ }\href {\doibase 10.1038/nature05881} {\bibfield
  {journal} {\bibinfo  {journal} {Nature}\ }\textbf {\bibinfo {volume} {447}},\
  \bibinfo {pages} {569} (\bibinfo {year} {2007})}\BibitemShut {NoStop}%
\bibitem [{\citenamefont {Parker}\ \emph {et~al.}(2010)\citenamefont {Parker},
  \citenamefont {Aynajian}, \citenamefont {{da Silva Neto}}, \citenamefont
  {Pushp}, \citenamefont {Ono}, \citenamefont {Wen}, \citenamefont {Xu},
  \citenamefont {Gu},\ and\ \citenamefont {Yazdani}}]{Parker2010}%
  \BibitemOpen
  \bibfield  {author} {\bibinfo {author} {\bibfnamefont {C.~V.}\ \bibnamefont
  {Parker}}, \bibinfo {author} {\bibfnamefont {P.}~\bibnamefont {Aynajian}},
  \bibinfo {author} {\bibfnamefont {E.~H.}\ \bibnamefont {{da Silva Neto}}},
  \bibinfo {author} {\bibfnamefont {A.}~\bibnamefont {Pushp}}, \bibinfo
  {author} {\bibfnamefont {S.}~\bibnamefont {Ono}}, \bibinfo {author}
  {\bibfnamefont {J.}~\bibnamefont {Wen}}, \bibinfo {author} {\bibfnamefont
  {Z.}~\bibnamefont {Xu}}, \bibinfo {author} {\bibfnamefont {G.}~\bibnamefont
  {Gu}}, \ and\ \bibinfo {author} {\bibfnamefont {A.}~\bibnamefont {Yazdani}},\
  }\href {http://dx.doi.org/10.1038/nature09597} {\bibfield  {journal}
  {\bibinfo  {journal} {Nature}\ }\textbf {\bibinfo {volume} {468}},\ \bibinfo
  {pages} {677} (\bibinfo {year} {2010})}\BibitemShut {NoStop}%
\bibitem [{\citenamefont {Wu}\ \emph {et~al.}(2011)\citenamefont {Wu} \emph
  {et~al.}}]{Wu2011}%
  \BibitemOpen
  \bibfield  {author} {\bibinfo {author} {\bibfnamefont {T.}~\bibnamefont {Wu}}
  \emph {et~al.},\ }\href {http://dx.doi.org/10.1038/nature10345
  http://10.0.4.14/nature10345
  https://www.nature.com/articles/nature10345{\#}supplementary-information}
  {\bibfield  {journal} {\bibinfo  {journal} {Nature}\ }\textbf {\bibinfo
  {volume} {477}},\ \bibinfo {pages} {191} (\bibinfo {year}
  {2011})}\BibitemShut {NoStop}%
\bibitem [{\citenamefont {Chang}\ \emph {et~al.}(2012)\citenamefont {Chang}
  \emph {et~al.}}]{Chang2012}%
  \BibitemOpen
  \bibfield  {author} {\bibinfo {author} {\bibfnamefont {J.}~\bibnamefont
  {Chang}} \emph {et~al.},\ }\href {\doibase 10.1038/nphys2456} {\bibfield
  {journal} {\bibinfo  {journal} {Nature Physics}\ }\textbf {\bibinfo {volume}
  {8}},\ \bibinfo {pages} {871} (\bibinfo {year} {2012})}\BibitemShut {NoStop}%
\bibitem [{\citenamefont {Blanco-Canosa}\ \emph {et~al.}(2014)\citenamefont
  {Blanco-Canosa} \emph {et~al.}}]{Blanco-Canosa2014}%
  \BibitemOpen
  \bibfield  {author} {\bibinfo {author} {\bibfnamefont {S.}~\bibnamefont
  {Blanco-Canosa}} \emph {et~al.},\ }\href {\doibase
  10.1103/PhysRevB.90.054513} {\bibfield  {journal} {\bibinfo  {journal}
  {Physical Review B}\ }\textbf {\bibinfo {volume} {90}},\ \bibinfo {pages}
  {054513} (\bibinfo {year} {2014})}\BibitemShut {NoStop}%
\bibitem [{\citenamefont {H{\"{u}}cker}\ \emph {et~al.}(2014)\citenamefont
  {H{\"{u}}cker} \emph {et~al.}}]{Huecker2014}%
  \BibitemOpen
  \bibfield  {author} {\bibinfo {author} {\bibfnamefont {M.}~\bibnamefont
  {H{\"{u}}cker}} \emph {et~al.},\ }\href {\doibase 10.1103/PhysRevB.90.054514}
  {\bibfield  {journal} {\bibinfo  {journal} {Physical Review B}\ }\textbf
  {\bibinfo {volume} {90}},\ \bibinfo {pages} {1} (\bibinfo {year}
  {2014})}\BibitemShut {NoStop}%
\bibitem [{\citenamefont {{da Silva Neto}}\ \emph {et~al.}(2014)\citenamefont
  {{da Silva Neto}} \emph {et~al.}}]{DaSilvaNeto2014}%
  \BibitemOpen
  \bibfield  {author} {\bibinfo {author} {\bibfnamefont {E.~H.}\ \bibnamefont
  {{da Silva Neto}}} \emph {et~al.},\ }\href {\doibase 10.1126/science.1243479}
  {\bibfield  {journal} {\bibinfo  {journal} {Science}\ }\textbf {\bibinfo
  {volume} {343}},\ \bibinfo {pages} {393} (\bibinfo {year} {2014})},\ \Eprint
  {http://arxiv.org/abs/1105.2508} {1105.2508} \BibitemShut {NoStop}%
\bibitem [{\citenamefont {Campi}\ \emph {et~al.}(2015)\citenamefont {Campi},
  \emph {et~al.}}]{Campi2015}%
  \BibitemOpen
  \bibfield  {author} {\bibinfo {author} {\bibfnamefont {G.}~\bibnamefont
  {Campi}}, ,  \emph {et~al.},\ }\href {\doibase 10.1038/nature14987}
  {\bibfield  {journal} {\bibinfo  {journal} {Nature}\ }\textbf {\bibinfo
  {volume} {525}},\ \bibinfo {pages} {359} (\bibinfo {year}
  {2015})}\BibitemShut {NoStop}%
\bibitem [{\citenamefont {Comin}\ \emph {et~al.}(2015)\citenamefont {Comin}
  \emph {et~al.}}]{Comin2015a}%
  \BibitemOpen
  \bibfield  {author} {\bibinfo {author} {\bibfnamefont {R.}~\bibnamefont
  {Comin}} \emph {et~al.},\ }\href {\doibase 10.1126/science.1258399}
  {\bibfield  {journal} {\bibinfo  {journal} {Science (New York, N.Y.)}\
  }\textbf {\bibinfo {volume} {347}},\ \bibinfo {pages} {1335} (\bibinfo {year}
  {2015})}\BibitemShut {NoStop}%
\bibitem [{\citenamefont {Wu}\ \emph {et~al.}(2017)\citenamefont {Wu},
  \citenamefont {Bollinger}, \citenamefont {He},\ and\ \citenamefont
  {Bo{\v{z}}ovi{\'c}}}]{Wu2017}%
  \BibitemOpen
  \bibfield  {author} {\bibinfo {author} {\bibfnamefont {J.}~\bibnamefont
  {Wu}}, \bibinfo {author} {\bibfnamefont {A.~T.}\ \bibnamefont {Bollinger}},
  \bibinfo {author} {\bibfnamefont {X.}~\bibnamefont {He}}, \ and\ \bibinfo
  {author} {\bibfnamefont {I.}~\bibnamefont {Bo{\v{z}}ovi{\'c}}},\ }\href
  {http://dx.doi.org/10.1038/nature23290} {\bibfield  {journal} {\bibinfo
  {journal} {Nature}\ }\textbf {\bibinfo {volume} {547}},\ \bibinfo {pages}
  {432} (\bibinfo {year} {2017})}\BibitemShut {NoStop}%
\bibitem [{\citenamefont {Tabis}\ \emph {et~al.}(2017)\citenamefont {Tabis}
  \emph {et~al.}}]{Tabis2017}%
  \BibitemOpen
  \bibfield  {author} {\bibinfo {author} {\bibfnamefont {W.}~\bibnamefont
  {Tabis}} \emph {et~al.},\ }\href {\doibase 10.1103/PhysRevB.96.134510}
  {\bibfield  {journal} {\bibinfo  {journal} {Phys. Rev. B}\ }\textbf {\bibinfo
  {volume} {96}},\ \bibinfo {pages} {134510} (\bibinfo {year}
  {2017})}\BibitemShut {NoStop}%
\bibitem [{\citenamefont {Fradkin}\ and\ \citenamefont
  {Kivelson}(2012)}]{Fradkin2012}%
  \BibitemOpen
  \bibfield  {author} {\bibinfo {author} {\bibfnamefont {E.}~\bibnamefont
  {Fradkin}}\ and\ \bibinfo {author} {\bibfnamefont {S.~A.}\ \bibnamefont
  {Kivelson}},\ }\href {\doibase 10.1038/nphys2498} {\bibfield  {journal}
  {\bibinfo  {journal} {Nature Physics}\ }\textbf {\bibinfo {volume} {8}},\
  \bibinfo {pages} {864} (\bibinfo {year} {2012})}\BibitemShut {NoStop}%
\bibitem [{\citenamefont {Plakida}\ \emph {et~al.}(2003)\citenamefont
  {Plakida}, \citenamefont {Anton}, \citenamefont {Adam},\ and\ \citenamefont
  {Adam}}]{Plakida2003}%
  \BibitemOpen
  \bibfield  {author} {\bibinfo {author} {\bibfnamefont {N.~M.}\ \bibnamefont
  {Plakida}}, \bibinfo {author} {\bibfnamefont {L.}~\bibnamefont {Anton}},
  \bibinfo {author} {\bibfnamefont {S.}~\bibnamefont {Adam}}, \ and\ \bibinfo
  {author} {\bibfnamefont {G.}~\bibnamefont {Adam}},\ }\href {\doibase
  10.1134/1.1608998} {\bibfield  {journal} {\bibinfo  {journal} {Journal of
  Experimental and Theoretical Physics}\ }\textbf {\bibinfo {volume} {97}},\
  \bibinfo {pages} {331} (\bibinfo {year} {2003})}\BibitemShut {NoStop}%
\bibitem [{\citenamefont {Plakida}\ and\ \citenamefont
  {Oudovenko}(2016)}]{Plakida2016}%
  \BibitemOpen
  \bibfield  {author} {\bibinfo {author} {\bibfnamefont {N.~M.}\ \bibnamefont
  {Plakida}}\ and\ \bibinfo {author} {\bibfnamefont {V.~S.}\ \bibnamefont
  {Oudovenko}},\ }\href {\doibase 10.1007/s10948-016-3379-4} {\bibfield
  {journal} {\bibinfo  {journal} {Journal of Superconductivity and Novel
  Magnetism}\ }\textbf {\bibinfo {volume} {29}},\ \bibinfo {pages} {1037}
  (\bibinfo {year} {2016})}\BibitemShut {NoStop}%
\bibitem [{\citenamefont {Vojta}(2002)}]{Vojta2002}%
  \BibitemOpen
  \bibfield  {author} {\bibinfo {author} {\bibfnamefont {M.}~\bibnamefont
  {Vojta}},\ }\href {\doibase 10.1103/PhysRevB.66.104505} {\bibfield  {journal}
  {\bibinfo  {journal} {Phys. Rev. B}\ }\textbf {\bibinfo {volume} {66}},\
  \bibinfo {pages} {104505} (\bibinfo {year} {2002})}\BibitemShut {NoStop}%
\bibitem [{\citenamefont {Ortix}\ \emph {et~al.}(2008)\citenamefont {Ortix},
  \citenamefont {Lorenzana},\ and\ \citenamefont {Di~Castro}}]{DiCastro2008}%
  \BibitemOpen
  \bibfield  {author} {\bibinfo {author} {\bibfnamefont {C.}~\bibnamefont
  {Ortix}}, \bibinfo {author} {\bibfnamefont {J.}~\bibnamefont {Lorenzana}}, \
  and\ \bibinfo {author} {\bibfnamefont {C.}~\bibnamefont {Di~Castro}},\ }\href
  {\doibase 10.1103/PhysRevLett.100.246402} {\bibfield  {journal} {\bibinfo
  {journal} {Phys. Rev. Lett.}\ }\textbf {\bibinfo {volume} {100}},\ \bibinfo
  {pages} {246402} (\bibinfo {year} {2008})}\BibitemShut {NoStop}%
\bibitem [{\citenamefont {Nie}\ \emph {et~al.}(2017)\citenamefont {Nie},
  \citenamefont {Maharaj}, \citenamefont {Fradkin},\ and\ \citenamefont
  {Kivelson}}]{Kivelson2017}%
  \BibitemOpen
  \bibfield  {author} {\bibinfo {author} {\bibfnamefont {L.}~\bibnamefont
  {Nie}}, \bibinfo {author} {\bibfnamefont {A.~V.}\ \bibnamefont {Maharaj}},
  \bibinfo {author} {\bibfnamefont {E.}~\bibnamefont {Fradkin}}, \ and\
  \bibinfo {author} {\bibfnamefont {S.~A.}\ \bibnamefont {Kivelson}},\ }\href
  {\doibase 10.1103/PhysRevB.96.085142} {\bibfield  {journal} {\bibinfo
  {journal} {Phys. Rev. B}\ }\textbf {\bibinfo {volume} {96}},\ \bibinfo
  {pages} {085142} (\bibinfo {year} {2017})}\BibitemShut {NoStop}%
\bibitem [{\citenamefont {Okamoto}\ \emph {et~al.}(2010)\citenamefont
  {Okamoto}, \citenamefont {S\'en\'echal}, \citenamefont {Civelli},\ and\
  \citenamefont {Tremblay}}]{AMarie2010}%
  \BibitemOpen
  \bibfield  {author} {\bibinfo {author} {\bibfnamefont {S.}~\bibnamefont
  {Okamoto}}, \bibinfo {author} {\bibfnamefont {D.}~\bibnamefont
  {S\'en\'echal}}, \bibinfo {author} {\bibfnamefont {M.}~\bibnamefont
  {Civelli}}, \ and\ \bibinfo {author} {\bibfnamefont {A.-M.~S.}\ \bibnamefont
  {Tremblay}},\ }\href {\doibase 10.1103/PhysRevB.82.180511} {\bibfield
  {journal} {\bibinfo  {journal} {Phys. Rev. B}\ }\textbf {\bibinfo {volume}
  {82}},\ \bibinfo {pages} {180511} (\bibinfo {year} {2010})}\BibitemShut
  {NoStop}%
\bibitem [{\citenamefont {Kharkov}\ and\ \citenamefont
  {Sushkov}(2016)}]{Kharkov2016}%
  \BibitemOpen
  \bibfield  {author} {\bibinfo {author} {\bibfnamefont {Y.~A.}\ \bibnamefont
  {Kharkov}}\ and\ \bibinfo {author} {\bibfnamefont {O.~P.}\ \bibnamefont
  {Sushkov}},\ }\href {http://dx.doi.org/10.1038/srep34551
  http://10.0.4.14/srep34551
  https://www.nature.com/articles/srep34551{\#}supplementary-information}
  {\bibfield  {journal} {\bibinfo  {journal} {Scientific Reports}\ }\textbf
  {\bibinfo {volume} {6}},\ \bibinfo {pages} {34551} (\bibinfo {year}
  {2016})}\BibitemShut {NoStop}%
\bibitem [{\citenamefont {Lang}\ \emph {et~al.}(2002)\citenamefont {Lang} \emph
  {et~al.}}]{Lang2002}%
  \BibitemOpen
  \bibfield  {author} {\bibinfo {author} {\bibfnamefont {K.~M.}\ \bibnamefont
  {Lang}} \emph {et~al.},\ }\href {\doibase 10.1038/415412a} {\bibfield
  {journal} {\bibinfo  {journal} {Nature}\ }\textbf {\bibinfo {volume} {415}},\
  \bibinfo {pages} {412} (\bibinfo {year} {2002})}\BibitemShut {NoStop}%
\bibitem [{\citenamefont {Hanaguri}\ \emph {et~al.}(2004)\citenamefont
  {Hanaguri} \emph {et~al.}}]{Hanaguri2004}%
  \BibitemOpen
  \bibfield  {author} {\bibinfo {author} {\bibfnamefont {T.}~\bibnamefont
  {Hanaguri}} \emph {et~al.},\ }\href {\doibase 10.1038/nature02861} {\bibfield
   {journal} {\bibinfo  {journal} {Nature}\ }\textbf {\bibinfo {volume}
  {430}},\ \bibinfo {pages} {1001} (\bibinfo {year} {2004})}\BibitemShut
  {NoStop}%
\bibitem [{\citenamefont {Cahn}\ and\ \citenamefont
  {Hilliard}(1958)}]{Cahn1958}%
  \BibitemOpen
  \bibfield  {author} {\bibinfo {author} {\bibfnamefont {J.~W.}\ \bibnamefont
  {Cahn}}\ and\ \bibinfo {author} {\bibfnamefont {J.~E.}\ \bibnamefont
  {Hilliard}},\ }\href {\doibase 10.1063/1.1744102} {\bibfield  {journal}
  {\bibinfo  {journal} {J. Chem. Phys.}\ }\textbf {\bibinfo {volume} {28}},\
  \bibinfo {pages} {258} (\bibinfo {year} {1958})}\BibitemShut {NoStop}%
\bibitem [{\citenamefont {deMello}\ \emph {et~al.}(2009)\citenamefont
  {deMello}, \citenamefont {Kasal},\ and\ \citenamefont
  {Passos}}]{deMello2009}%
  \BibitemOpen
  \bibfield  {author} {\bibinfo {author} {\bibfnamefont {E.}~\bibnamefont
  {deMello}}, \bibinfo {author} {\bibfnamefont {R.}~\bibnamefont {Kasal}}, \
  and\ \bibinfo {author} {\bibfnamefont {C.}~\bibnamefont {Passos}},\ }\href
  {http://stacks.iop.org/0953-8984/21/i=23/a=235701} {\bibfield  {journal}
  {\bibinfo  {journal} {J. Phys.: Condens. Matter}\ }\textbf {\bibinfo {volume}
  {21}},\ \bibinfo {pages} {235701} (\bibinfo {year} {2009})}\BibitemShut
  {NoStop}%
\bibitem [{\citenamefont {deMello}(2012)}]{DeMello2012}%
  \BibitemOpen
  \bibfield  {author} {\bibinfo {author} {\bibfnamefont {E.}~\bibnamefont
  {deMello}},\ }\href {\doibase 10.1209/0295-5075/99/37003} {\bibfield
  {journal} {\bibinfo  {journal} {Europhys. Lett.}\ }\textbf {\bibinfo {volume}
  {99}},\ \bibinfo {pages} {37003} (\bibinfo {year} {2012})}\BibitemShut
  {NoStop}%
\bibitem [{\citenamefont {deMello}\ and\ \citenamefont
  {Kasal}(2012)}]{deMelloKasal2012}%
  \BibitemOpen
  \bibfield  {author} {\bibinfo {author} {\bibfnamefont {E.}~\bibnamefont
  {deMello}}\ and\ \bibinfo {author} {\bibfnamefont {R.}~\bibnamefont
  {Kasal}},\ }\href
  {http://www.sciencedirect.com/science/article/pii/S092145341100503X}
  {\bibfield  {journal} {\bibinfo  {journal} {Physica C: Superconductivity}\
  }\textbf {\bibinfo {volume} {472}},\ \bibinfo {pages} {60} (\bibinfo {year}
  {2012})}\BibitemShut {NoStop}%
\bibitem [{\citenamefont {de~Mello}\ and\ \citenamefont
  {Sonier}(2014)}]{DeMello2014}%
  \BibitemOpen
  \bibfield  {author} {\bibinfo {author} {\bibfnamefont {E.}~\bibnamefont
  {de~Mello}}\ and\ \bibinfo {author} {\bibfnamefont {J.}~\bibnamefont
  {Sonier}},\ }\href {http://stacks.iop.org/0953-8984/26/i=49/a=492201}
  {\bibfield  {journal} {\bibinfo  {journal} {J. Phys.: Condens. Matter}\
  }\textbf {\bibinfo {volume} {26}},\ \bibinfo {pages} {492201} (\bibinfo
  {year} {2014})}\BibitemShut {NoStop}%
\bibitem [{\citenamefont {deMello}\ and\ \citenamefont
  {Sonier}(2017)}]{Mello2017}%
  \BibitemOpen
  \bibfield  {author} {\bibinfo {author} {\bibfnamefont {E.}~\bibnamefont
  {deMello}}\ and\ \bibinfo {author} {\bibfnamefont {J.}~\bibnamefont
  {Sonier}},\ }\href {\doibase 10.1103/PhysRevB.95.184520} {\bibfield
  {journal} {\bibinfo  {journal} {Phys. Rev. B}\ }\textbf {\bibinfo {volume}
  {95}},\ \bibinfo {pages} {184520} (\bibinfo {year} {2017})}\BibitemShut
  {NoStop}%
\bibitem [{\citenamefont {deMello}\ and\ \citenamefont
  {Filho}(2005)}]{Otton2005}%
  \BibitemOpen
  \bibfield  {author} {\bibinfo {author} {\bibfnamefont {E.}~\bibnamefont
  {deMello}}\ and\ \bibinfo {author} {\bibfnamefont {O.~S.}\ \bibnamefont
  {Filho}},\ }\href {\doibase http://dx.doi.org/10.1016/j.physa.2004.08.076}
  {\bibfield  {journal} {\bibinfo  {journal} {Physica A}\ }\textbf {\bibinfo
  {volume} {347}},\ \bibinfo {pages} {429 } (\bibinfo {year}
  {2005})}\BibitemShut {NoStop}%
\bibitem [{\citenamefont {Bray}(1994)}]{Bray1994}%
  \BibitemOpen
  \bibfield  {author} {\bibinfo {author} {\bibfnamefont {A.}~\bibnamefont
  {Bray}},\ }\href {\doibase 10.1080/00018739400101505} {\bibfield  {journal}
  {\bibinfo  {journal} {Adv. Phys.}\ }\textbf {\bibinfo {volume} {43}},\
  \bibinfo {pages} {357} (\bibinfo {year} {1994})}\BibitemShut {NoStop}%
\bibitem [{\citenamefont {Tranquada}\ \emph {et~al.}(1995)\citenamefont
  {Tranquada}, \citenamefont {Sternlieb}, \citenamefont {Axe}, \citenamefont
  {Nakamura},\ and\ \citenamefont {Uchida}}]{Tranquada1995a}%
  \BibitemOpen
  \bibfield  {author} {\bibinfo {author} {\bibfnamefont {J.~M.}\ \bibnamefont
  {Tranquada}}, \bibinfo {author} {\bibfnamefont {B.~J.}\ \bibnamefont
  {Sternlieb}}, \bibinfo {author} {\bibfnamefont {J.~D.}\ \bibnamefont {Axe}},
  \bibinfo {author} {\bibfnamefont {Y.}~\bibnamefont {Nakamura}}, \ and\
  \bibinfo {author} {\bibfnamefont {S.}~\bibnamefont {Uchida}},\ }\href
  {\doibase 10.1038/375561a0} {\bibfield  {journal} {\bibinfo  {journal}
  {Nature}\ }\textbf {\bibinfo {volume} {375}},\ \bibinfo {pages} {561}
  (\bibinfo {year} {1995})}\BibitemShut {NoStop}%
\bibitem [{\citenamefont {Thampy}\ \emph {et~al.}(2017)\citenamefont {Thampy},
  \citenamefont {Chen}, \citenamefont {Cao}, \citenamefont {Mazzoli},
  \citenamefont {Barbour}, \citenamefont {Hu}, \citenamefont {Miao},
  \citenamefont {Fabbris}, \citenamefont {Zhong}, \citenamefont {Gu},
  \citenamefont {Tranquada}, \citenamefont {Robinson}, \citenamefont
  {Wilkins},\ and\ \citenamefont {Dean}}]{Thampy2017}%
  \BibitemOpen
  \bibfield  {author} {\bibinfo {author} {\bibfnamefont {V.}~\bibnamefont
  {Thampy}}, \bibinfo {author} {\bibfnamefont {X.~M.}\ \bibnamefont {Chen}},
  \bibinfo {author} {\bibfnamefont {Y.}~\bibnamefont {Cao}}, \bibinfo {author}
  {\bibfnamefont {C.}~\bibnamefont {Mazzoli}}, \bibinfo {author} {\bibfnamefont
  {A.~M.}\ \bibnamefont {Barbour}}, \bibinfo {author} {\bibfnamefont
  {W.}~\bibnamefont {Hu}}, \bibinfo {author} {\bibfnamefont {H.}~\bibnamefont
  {Miao}}, \bibinfo {author} {\bibfnamefont {G.}~\bibnamefont {Fabbris}},
  \bibinfo {author} {\bibfnamefont {R.~D.}\ \bibnamefont {Zhong}}, \bibinfo
  {author} {\bibfnamefont {G.~D.}\ \bibnamefont {Gu}}, \bibinfo {author}
  {\bibfnamefont {J.~M.}\ \bibnamefont {Tranquada}}, \bibinfo {author}
  {\bibfnamefont {I.~K.}\ \bibnamefont {Robinson}}, \bibinfo {author}
  {\bibfnamefont {S.~B.}\ \bibnamefont {Wilkins}}, \ and\ \bibinfo {author}
  {\bibfnamefont {M.~P.~M.}\ \bibnamefont {Dean}},\ }\href {\doibase
  10.1103/PhysRevB.95.241111} {\bibfield  {journal} {\bibinfo  {journal} {Phys.
  Rev. B}\ }\textbf {\bibinfo {volume} {95}},\ \bibinfo {pages} {241111}
  (\bibinfo {year} {2017})}\BibitemShut {NoStop}%
\bibitem [{\citenamefont {Peng}\ \emph {et~al.}(2018)\citenamefont {Peng} \emph
  {et~al.}}]{CDWover2018}%
  \BibitemOpen
  \bibfield  {author} {\bibinfo {author} {\bibfnamefont {Y.~Y.}\ \bibnamefont
  {Peng}} \emph {et~al.},\ }\href {\doibase 10.1038/s41563-018-0108-3}
  {\bibfield  {journal} {\bibinfo  {journal} {Nature Materials}\ }\textbf
  {\bibinfo {volume} {17}},\ \bibinfo {pages} {697} (\bibinfo {year}
  {2018})}\BibitemShut {NoStop}%
\bibitem [{\citenamefont {Abbamonte}(2006)}]{Abbamonte2006}%
  \BibitemOpen
  \bibfield  {author} {\bibinfo {author} {\bibfnamefont {P.}~\bibnamefont
  {Abbamonte}},\ }\href {\doibase 10.1103/PhysRevB.74.195113} {\bibfield
  {journal} {\bibinfo  {journal} {Phys. Rev. B}\ }\textbf {\bibinfo {volume}
  {74}},\ \bibinfo {pages} {195113} (\bibinfo {year} {2006})}\BibitemShut
  {NoStop}%
\bibitem [{\citenamefont {Bendele}\ \emph {et~al.}(2017)\citenamefont {Bendele}
  \emph {et~al.}}]{Isotop2017}%
  \BibitemOpen
  \bibfield  {author} {\bibinfo {author} {\bibfnamefont {M.}~\bibnamefont
  {Bendele}} \emph {et~al.},\ }\href {\doibase 10.1103/PhysRevB.95.014514}
  {\bibfield  {journal} {\bibinfo  {journal} {Phys. Rev. B}\ }\textbf {\bibinfo
  {volume} {95}},\ \bibinfo {pages} {014514} (\bibinfo {year}
  {2017})}\BibitemShut {NoStop}%
\bibitem [{\citenamefont {Imry}\ \emph {et~al.}(2012)\citenamefont {Imry},
  \citenamefont {Strongin},\ and\ \citenamefont {Homes}}]{Imry2012}%
  \BibitemOpen
  \bibfield  {author} {\bibinfo {author} {\bibfnamefont {Y.}~\bibnamefont
  {Imry}}, \bibinfo {author} {\bibfnamefont {M.}~\bibnamefont {Strongin}}, \
  and\ \bibinfo {author} {\bibfnamefont {C.~C.}\ \bibnamefont {Homes}},\ }\href
  {\doibase 10.1103/PhysRevLett.109.067003} {\bibfield  {journal} {\bibinfo
  {journal} {Phys. Rev. Lett.}\ }\textbf {\bibinfo {volume} {109}},\ \bibinfo
  {pages} {067003} (\bibinfo {year} {2012})}\BibitemShut {NoStop}%
\bibitem [{\citenamefont {Carlson}\ \emph {et~al.}()\citenamefont {Carlson},
  \citenamefont {Emery}, \citenamefont {Kivelson},\ and\ \citenamefont
  {Orgad}}]{ECarlson2002}%
  \BibitemOpen
  \bibfield  {author} {\bibinfo {author} {\bibfnamefont {E.~W.}\ \bibnamefont
  {Carlson}}, \bibinfo {author} {\bibfnamefont {V.~J.}\ \bibnamefont {Emery}},
  \bibinfo {author} {\bibfnamefont {S.~A.}\ \bibnamefont {Kivelson}}, \ and\
  \bibinfo {author} {\bibfnamefont {D.}~\bibnamefont {Orgad}},\ }\href
  {http://lanl.arxiv.org/abs/0206217v1} {\ }\Eprint
  {http://arxiv.org/abs/cond-mat.supr-con/0206217v1}
  {cond-mat.supr-con/0206217v1} \BibitemShut {NoStop}%
\bibitem [{\citenamefont {Ando}\ \emph {et~al.}(2004)\citenamefont {Ando},
  \citenamefont {Komiya}, \citenamefont {Segawa}, \citenamefont {Ono},\ and\
  \citenamefont {Kurita}}]{Ando2004}%
  \BibitemOpen
  \bibfield  {author} {\bibinfo {author} {\bibfnamefont {Y.}~\bibnamefont
  {Ando}}, \bibinfo {author} {\bibfnamefont {S.}~\bibnamefont {Komiya}},
  \bibinfo {author} {\bibfnamefont {K.}~\bibnamefont {Segawa}}, \bibinfo
  {author} {\bibfnamefont {S.}~\bibnamefont {Ono}}, \ and\ \bibinfo {author}
  {\bibfnamefont {Y.}~\bibnamefont {Kurita}},\ }\href {\doibase
  10.1103/PhysRevLett.93.267001} {\bibfield  {journal} {\bibinfo  {journal}
  {Phys. Rev. Lett.}\ }\textbf {\bibinfo {volume} {93}},\ \bibinfo {pages}
  {267001} (\bibinfo {year} {2004})}\BibitemShut {NoStop}%
\bibitem [{\citenamefont {Lee}\ \emph {et~al.}(2007)\citenamefont {Lee} \emph
  {et~al.}}]{Lee2007}%
  \BibitemOpen
  \bibfield  {author} {\bibinfo {author} {\bibfnamefont {W.~S.}\ \bibnamefont
  {Lee}} \emph {et~al.},\ }\href {\doibase 10.1038/nature06219} {\bibfield
  {journal} {\bibinfo  {journal} {Nature}\ }\textbf {\bibinfo {volume} {450}},\
  \bibinfo {pages} {81} (\bibinfo {year} {2007})}\BibitemShut {NoStop}%
\bibitem [{\citenamefont {Landau}\ and\ \citenamefont {Lifchitz}(1966)}]{LMQ}%
  \BibitemOpen
  \bibfield  {author} {\bibinfo {author} {\bibfnamefont {L.}~\bibnamefont
  {Landau}}\ and\ \bibinfo {author} {\bibfnamefont {E.}~\bibnamefont
  {Lifchitz}},\ }\href@noop {} {\emph {\bibinfo {title} {M\'ecanique
  Quantique}}}\ (\bibinfo  {publisher} {\'Editions Mir, Moscow, Russie},\
  \bibinfo {year} {1966})\BibitemShut {NoStop}%
\bibitem [{\citenamefont {Kanigel}\ \emph {et~al.}(2008)\citenamefont {Kanigel}
  \emph {et~al.}}]{Kanigel2008}%
  \BibitemOpen
  \bibfield  {author} {\bibinfo {author} {\bibfnamefont {A.}~\bibnamefont
  {Kanigel}} \emph {et~al.},\ }\href {\doibase 10.1103/PhysRevLett.101.137002}
  {\bibfield  {journal} {\bibinfo  {journal} {Phys. Rev. Lett.}\ }\textbf
  {\bibinfo {volume} {101}},\ \bibinfo {pages} {137002} (\bibinfo {year}
  {2008})}\BibitemShut {NoStop}%
\bibitem [{\citenamefont {Dubroka}\ \emph {et~al.}(2011)\citenamefont {Dubroka}
  \emph {et~al.}}]{Dubroka2011}%
  \BibitemOpen
  \bibfield  {author} {\bibinfo {author} {\bibfnamefont {A.}~\bibnamefont
  {Dubroka}} \emph {et~al.},\ }\href {\doibase 10.1103/PhysRevLett.106.047006}
  {\bibfield  {journal} {\bibinfo  {journal} {Phys. Rev. Lett.}\ }\textbf
  {\bibinfo {volume} {106}},\ \bibinfo {pages} {1} (\bibinfo {year}
  {2011})}\BibitemShut {NoStop}%
\bibitem [{\citenamefont {Bilbro}\ \emph {et~al.}(2011)\citenamefont {Bilbro},
  \citenamefont {Aguilar}, \citenamefont {Logvenov}, \citenamefont {Pelleg},
  \citenamefont {Boz̆ovi{\'{c}}},\ and\ \citenamefont
  {Armitage}}]{Bilbro2011}%
  \BibitemOpen
  \bibfield  {author} {\bibinfo {author} {\bibfnamefont {L.~S.}\ \bibnamefont
  {Bilbro}}, \bibinfo {author} {\bibfnamefont {R.~V.}\ \bibnamefont {Aguilar}},
  \bibinfo {author} {\bibfnamefont {G.}~\bibnamefont {Logvenov}}, \bibinfo
  {author} {\bibfnamefont {O.}~\bibnamefont {Pelleg}}, \bibinfo {author}
  {\bibfnamefont {I.}~\bibnamefont {Boz̆ovi{\'{c}}}}, \ and\ \bibinfo {author}
  {\bibfnamefont {N.~P.}\ \bibnamefont {Armitage}},\ }\href
  {http://dx.doi.org/10.1038/nphys1912 http://10.0.4.14/nphys1912
  https://www.nature.com/articles/nphys1912{\#}supplementary-information}
  {\bibfield  {journal} {\bibinfo  {journal} {Nature Physics}\ }\textbf
  {\bibinfo {volume} {7}},\ \bibinfo {pages} {298} (\bibinfo {year}
  {2011})}\BibitemShut {NoStop}%
\bibitem [{\citenamefont {deMello}(2020)}]{Mello2020b}%
  \BibitemOpen
  \bibfield  {author} {\bibinfo {author} {\bibfnamefont {E.}~\bibnamefont
  {deMello}},\ }\href {http://lanl.arxiv.org/abs/2001.07249} {} (\bibinfo
  {year} {2020}),\ \Eprint {http://arxiv.org/abs/arXiv:2001.07249}
  {arXiv:2001.07249} \BibitemShut {NoStop}%
\bibitem [{\citenamefont {de~Mello}(2020)}]{Mello2020c}%
  \BibitemOpen
  \bibfield  {author} {\bibinfo {author} {\bibfnamefont {E.~V.~L.}\
  \bibnamefont {de~Mello}},\ }\href {https://doi.org/10.1088/1361-648X/ab9407}
  {\bibfield  {journal} {\bibinfo  {journal} {Journal of Physics: Condensed
  Matter}\ } (\bibinfo {year} {2020})}\BibitemShut {NoStop}%
\end{thebibliography}
\end{document}